\documentclass[10pt,letterpaper,twocolumn]{article} 

\usepackage{ol2}
\usepackage[draft]{hyperref}
\usepackage{amsmath}
\usepackage[english,french]{babel}

\begin{document}

\twocolumn[ 

\title{Experimental wavelength division multiplexed photon pair distribution}


\author{Joe Ghalbouni, Imad Agha, Robert Frey, Eleni Diamanti and Isabelle Zaquine}

\address{Institut T\'el\'ecom/T\'el\'ecom Paristech, CNRS-LTCI 46 rue Barrault, 75013 Paris, France
\\
$^*$Corresponding author: isabelle.zaquine@telecom-paristech.fr
}

\begin{abstract}We have experimentally implemented the distribution of photon pairs produced by spontaneous parametric down conversion through telecom dense wavelength division multiplexing filters. Using the measured counts and coincidences between symmetric channels, we evaluate the maximum fringe visibility that can be obtained with polarization entangled photons and compare different filter technologies.
\end{abstract}


 ] 
\noindent In order to be truly useful for technological applications, future quantum communication networks will require a large number of high quality entangled photon pair sources. An attractive way to reduce the associated cost is to use the wide spectrum produced by spontaneous parametric down conversion (SPDC) in combination with standard off-the-shelf dense wavelength division multiplexing (DWDM) filters to distribute non-degenerate photon pairs to a large number of users. It is then an important task to test photon pair sources in a WDM environment with various types of filters, in order to unveil filter characteristics that are necessary to obtain sources featuring high visibility while maintaining a useful brightness. Among previous works based on the idea of DWDM distribution of photon pairs~\cite{Lim,Oh,Her,Sau}, Lim et al \cite{Lim} performed a proof-of-principle experiment using a dichroic mirror to separate the two photons of the pair and tunable filters on each channel to simulate the demultiplexing operation. Furthermore, a wavelength selective switch, based on arrayed waveguide grating technology, was tested in \cite{Oh}. Frequency dependent losses were taken into account to explain the experimental coincidence probabilities, however the system required a different tuning of the pump for distribution over the various channel pairs, hence it could not be simultaneously multi-user in that configuration. In this work, we compare various DWDM filters with respect to the quality of a photon pair source with the ultimate goal of determining the filter performances required for quantum communication applications.

The WDM photon pair distribution device is based on the energy conservation condition of spontaneous parametric down conversion : if $\omega_p$ is the pump frequency, then the frequencies of the signal and idler photons $\omega_s$ and $\omega_i$ are symmetric with respect to $\omega_p/2$. The idea then is to tune the degeneracy frequency $\omega_p/2$ to the central frequency of the filter, so that the signal and idler photons can be transmitted by symmetric channel pairs. The users that will receive the two photons of an entangled pair can be determined by using a standard telecom switching device. If we wanted each user to receive a given channel permanently, then distributing entanglement, for instance, to Alice ($\omega_A$) and Bob ($\omega_B$), would mean tuning the pump frequency to $\omega_A+\omega_B$.

Commercial optical demultiplexing filters are based on three main technologies \cite{Dutta2003}: (a) dielectric thin-film filters (DTF), consisting of Fabry-Perot cavities and quarter wavelength layers; these transmission bandpass filters are cascaded in order to separate the different channels, (b) arrayed-waveguide gratings (AWG) that are planar lightwave circuits and are based on multi-beam interference; the constructive interference condition in the output coupling region is satisfied at different focal points for the different wavelengths that can then be coupled into different output waveguides, (c) diffraction gratings (DG), which are free space diffraction gratings used in combination with imaging optics; due to the wavelength-dependent diffraction angle, diffracted beams with different wavelengths, are focused into different locations and then coupled into output fibers. For all technologies, the shape of the transmission curves can either be close to a Gaussian one or have a flat-top structure.

When such filters are used in a photon pair distribution device, the main performance limitation is expected to arise from the loss of coherence induced by various imperfections of the demultiplexer that splits the photon pairs. More specifically, the relevant filter parameters are the insertion loss, the stability and uniformity of the channel bandwidth, the precision of the channel center wavelength, as well as chromatic dispersion and group delay, characteristics that can greatly vary for different filter technologies. These parameters have a direct effect on the quality of an entangled photon source, which is typically evaluated using the visibility and the brightness as the most important figures of merit.

\begin{figure}[htb]
\centerline{\includegraphics[width=7.5cm]{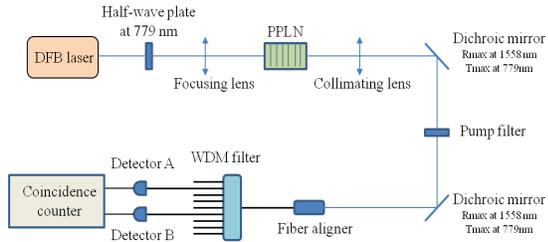}}
\caption{Experimental setup.}\label{Setup}
\end{figure}

The experimental setup is depicted in Fig \ref{Setup}. The pump is generated by a continuous-wave distributed feedback (DFB) laser at 779 nm, with a 20 mW power, and is focused in a 2 cm long MgO-doped periodically poled lithium niobate (PPLN) bulk crystal. The phase-matching condition is satisfied by adjusting the crystal temperature (typically $65^o$C) and pump, signal and idler waves have the same polarization in order to use the highest nonlinear coefficient of the crystal $d_{33}$. A halfwave plate is used to change the pump polarization, in order to control the efficiency of the nonlinear process and consequently the pair generation probability. After the crystal, all pump photons are filtered using dichroic mirrors and colored glass filters. The photon pairs are then coupled into a single-mode fiber and go through the DWDM filter, which features a channel separation of $100$ GHz and a channel width of $100$ GHz. In order to measure single counts at the signal and idler channels and coincidences, InGaAs avalanche diode single-photon detectors are connected either to two symmetric channels of the filter to obtain a deterministic splitting of the pairs or to the two outputs of a balanced fiber beamsplitter to obtain a statistical splitting. The detectors are triggered at $2$ MHz. Detector noise is $500$ counts/s for a $20$ ns gate and detector dead time is $10~\mu$s.

We have shown in previous work \cite{Smi1} that using only single count and coincidence measurements, in combination with detector noise count data, the maximum fringe visibility that can be obtained when entanglement is implemented with such a photon pair source can be predicted accurately. The calculation takes into account the visibility degradation due to double pair generation in SPDC, and examines the dependence of the visibility on the channel insertion losses and the shape of the filter transmission. In particular, we use single count, coincidence and noise count measurements performed with our setup to determine the ratio of true to accidental coincidences $P_{\text{TC}}/P_{\text{AC}}$, and thus evaluate the maximum attainable visibility $V_{\text{max}}$ of the source as follows \cite{Smi1}:
\begin{equation}\label{eq:V}
V_{\text{max}}=\frac{1}{1+2\frac{P_{\text{AC}}}{P_{\text{TC}}}}
\end{equation}

\noindent To characterize the influence of the DWDM filters, we measure their transmission using a wideband infrared source in order to determine the following integrals:
\begin{eqnarray}\label{eq:Omega2}
	I_1(\text{channel }i)  \!\! \!\!&=&  \!\!\!\! \!\! \!\! \int \!\! \frac{d\omega}{2\pi} \mathcal{T}(\omega-\omega_{F_i})\\
	I_2(\text{channel pair }ij) \!\! \!\!&=&   \!\!\!\!\!\! \!\!\int \!\! \frac{d\omega_s}{2\pi}  \mathcal{T}(\omega_s-\omega_{F})\nonumber \mathcal{T}(\omega_p-\omega_s-\omega_F)
\end{eqnarray}
where $\mathcal{T}(\omega)$ is the normalized intensity transmission of the filter, $\omega_{F_i}$ the center frequency of channel $i$ and $\omega_F$ the center frequency of all channel pairs. The single count probabilities measured at the output of channel $i$ are proportional to $I_1(\text{channel }i)$, while the coincidence probability between channel $i$ and channel $j$ is proportional to $I_2$, which is maximum when the center frequency of the filter is equal to the SPDC degeneracy frequency $\omega_p/2=\omega_F$. Using the values of $I_1$ and $I_2$ for the considered channel or channel pair, we can then determine the down conversion probability within the filter bandwidth or inband pair emission probability $p(I_1)$:
\begin{eqnarray}\label{eq:p}
\textrm{statistical splitting} \qquad p(I_1)&=&\frac{I_2}{2I_1}\frac{P_{\text{AC}}}{P_{\text{TC}}}\\
\textrm{deterministic splitting} \qquad p(I_1)&=&\frac{I_2}{I_1}\frac{P_{\text{AC}}}{P_{\text{TC}}}
\end{eqnarray}

In the deterministic splitting case, where signal and idler frequencies are different, true coincidences could be observed only when delaying the detection of one of the two filter channels. This is due to the demultiplexer technologies, where the wavelength separation is associated to a variable group delay.
This delay was measured for different channel pairs (see Table \ref{TDelay}) within the available delay range of the detectors (0-25 ns). For the AWG filter channel pair $21-27$, the delay was too large to be measured. For a given pump frequency, the diffraction grating technology gives rise to a constant delay for all channel pairs where as the dielectric thin film technology exhibits a delay varying from -2.5 to 22.5 ns. If the pump laser wavelength is fixed, then the distribution allows only fixed pairs of users. In this case, the delay can be compensated for each channel pair. If the pump laser can be tuned, in order to change the pairs of users that receive the two photons of one pair, a variable delay line should be used and a preliminary optimization of the coincidences would have to be implemented before the pair distribution can start.
\begin{table}
  \centering
  \caption{Group delays for different channel pairs for four different filters. The channel numbers correspond to the International Telecommunication Union (ITU) grid.}\label{TDelay}
\begin{tabular}{ccccc} \\ \hline
    Filter type & Channel numbers & Delay (ns)\\ \hline
 & 23-25 & 15 &\\
DTF (Flat-top)& 22-26 & 22.5 &\\
& 21-27 & -2.5 &\\\hline
AWG (Flat-top)& 23-25 & 12.5&\\
& 22-26 & 10&\\\hline
DG (Flat-top) & 23-25, 22-26, 21-27 & 10&\\\hline
DG (Gaussian) & 23-25, 22-26, 21-27 & 10& \\\hline
  \end{tabular}
\end{table}

\begin{figure}[htb]
\centerline{\includegraphics[width=7.5cm]{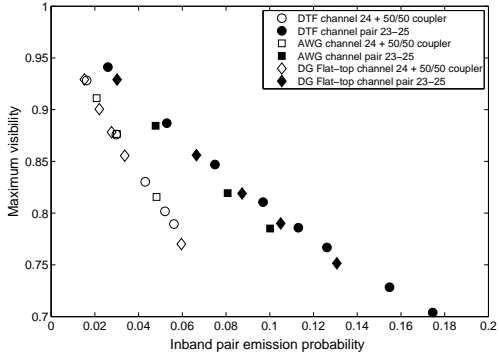}}
\caption{Maximum possible visibility as a function of inband pair emission probability $p(I_1)$ for the channel pair 23-25 and for the channel 24 using a 50/50 beamsplitter, with three different flat top filters.}\label{2325}
\end{figure}

\begin{figure}[htb]
\centerline{\includegraphics[width=7.5cm]{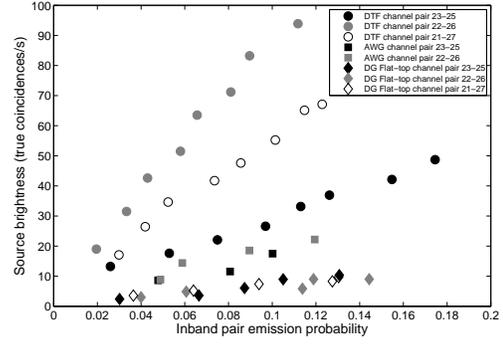}}
\caption{Number of true coincidences per second as a function of inband pair emission probability $p(I_1)$ for all channel pairs of the three flat top demultiplexers.}\label{allall}
\end{figure}

Fig. \ref{2325} shows the maximum visibility as a function of inband pair emission probability $p(I_1)$, using deterministic or statistical splitting, for the three different flat-top filters. The inband pair emission probability was varied between 0.01 and 0.18 to include a wide range of visibility values with a lower bound of $\sqrt{2}/2$, below which no entanglement is possible. Note that lower pair emission probabilities can yield higher visibility, but at the expense of impractical values for the source brightness, which corresponds to the number of true coincidences per second. As we observe in Fig. \ref{2325}, the deterministic splitting gives better results than the statistical one \cite{Smi2}. With respect to the various filter technologies, the AWG gives slightly lower visibility, but the difference is not large. The difference is clearly more significant when the source brightness is considered, as we can see in Fig. \ref{allall}. On the one hand, a large dispersion of this parameter is observed between different channel pairs of the dielectric thin film demultiplexer that cannot be explained solely by the insertion loss dispersion (the maximum channel transmission varies from 0.69 to 0.82). On the other hand, all the channel pairs of this demultiplexer give higher brightness than any of the arrayed-waveguide and diffraction grating demultiplexers.

\begin{figure}[htb]
\centerline{\includegraphics[width=7.5cm]{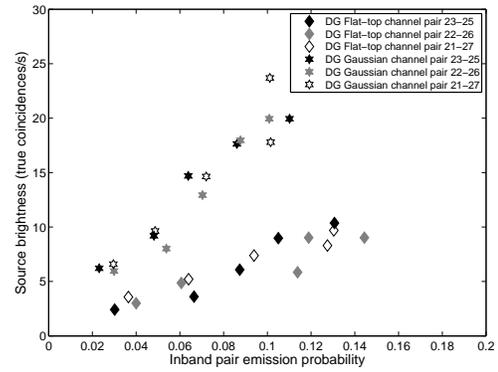}}
\caption{Number of true coincidences per second as a function of inband pair emission probability for the diffraction grating filter, comparing Gaussian and flat-top shapes for the channel pairs 23-25, 22-26, and 21-27.}\label{DGbri}
\end{figure}

Fig. \ref{DGbri} shows both the good uniformity of the brightness for the different channel pairs of the flat-top and Gaussian diffraction grating filters and the impact of the transmission curve shape. Although we have confirmed that the visibility of the Gaussian filter is only slightly lower than that of the flat-top, the corresponding brightness shows a very significant difference in favor of the Gaussian shape, due its higher transmission efficiency.



We have shown experimentally that commercial DWDM filters can be used to distribute photon pairs to a large number of users, taking advantage of the large bandwidth of spontaneous parametric down conversion in a PPLN crystal. We studied the impact of the demultiplexer technology on the important figures of merit of the entanglement multi-user distribution, and showed that, for a given visibility, the dielectric thin film technology gives large brightness but also large inter-channel dispersion, whereas a good uniformity is provided by the diffraction grating filters, at the expense of a brightness that is two to four times smaller. Such a DWDM photon pair distribution device can provide an economical solution for the development of quantum communication networks as long as the requirements on filter performance are satisfied, depending on the application. The trade-off between source visibility and brightness is clearly one of the main issues that have to be considered in such an optimization process.

\pagebreak

\section*{Informational Fourth Page}
In this section, please provide full versions of citations to assist reviewers and editors (OL publishes a short form of citations) or any other information that would aid the peer-review process.

\end{document}